# Overview of EXL-50 Research Progress and Future Plan

Yuejiang SHI[*], Yumin WANG, Bing LIU, Xianming SONG, Shaodong SONG, Xinchen JIANG, Dong GUO, Di LUO, Xiang GU, Tiantian SUN, Xianli HUANG, Zhi LI, Lili DONG, Xueyun WANG, Gang Yin, Mingyuan WANG, Wenjun Liu, Hanyue ZHAO, Huasheng XIE, Yong, Liu, Dongkai QI, Bo XING, Jiangbo DING, Chao WU, Lei LI, Haijun ZHANG, Yuanming YANG, Xin ZHAO, Enwu YANG, Wenwu LUO, Peihai ZHOU, Lei HAN, Qing ZHOU, Hanqing WANG, Jiaqi DONG, Baoshan YUAN, Y.-K.Martin. PENG, Minsheng LIU and the EXL-50 Team[a]

Hebei Key Laboratory of Compact Fusion, Langfang 065001, China

ENN Science and Technology Development Co., Ltd., Langfang 065001, China

Email: yjshi@ipp.ac.cn, liuminsheng@enn.cn

[a] See the appendix B of [35]

**Abstract**

XuanLong-50 (EXL-50) is the first medium-size spherical torus (ST) in China, with the toroidal field at major radius at 50 cm around 0.5T. CS-free and non-inductive current drive via electron cyclotron resonance heating (ECRH) was the main physics research issue for EXL-50. Discharges with plasma currents of 50 kA -180 kA were routinely obtained in EXL-50, with the current flattop sustained for up to or beyond 2 s. The current drive effectiveness on EXL-50 was as high as 1 A/W for low-density discharges using 28GHz ECRH alone for heating power less than 200 kW. The plasma current reached $I_p > 80$ kA for high-density ($>5 \times 10^{18}$ m$^{-2}$) discharges with 150 kW 28GHz ECRH. Higher performance discharge ($I_p$ of about 120 kA and core density of about $1\times 10^{19}$ m$^{-3}$) was achieved with 150 kW 50GHz ECRH. The plasma current in EXL-50 was mainly carried by the energetic electrons. Multi-fluid equilibrium model has been successfully applied to reconstruct the magnetic flux surface and the measured plasma parameters of the EXL-50's equilibrium. The physics mechanisms for the solenoid-free ECRH current drive and the energetic electrons has also been investigated. Preliminary experimental results show that 100 kW of lower hybrid current drive (LHCD) waves can drive 20 kA of plasma current. Several boron injection systems were installed and tested in EXL-50, including B$_2$H$_6$ gas puffing, boron powder injection, boron pellet injection. The research plan of EXL-50U, which is the upgrade machine of EXL-50, is also presented.

## 1. INTRODUCTION

The tokamak has been the most investigated and furthest advanced configuration among the magnetic confinement fusion systems. More recently, the spherical torus (ST) concept of aspect ratios around 1.5 [1-2] has been experimentally (START [3], NSTX [4], MAST [5], and Globus-M2 [6-7], ST-40[8]) tested to realize a substantially higher plasma beta compared to the tokamak of aspect ratios around 3, and is an attractive candidate for realizing a relatively compact fusion reactor. However, whether the central core of ST can survive the high dose of 14 MeV neutrons for a long enough time is one of the biggest challenges for the application of ST in future commercial D-T fusion reactor. On the other hand, the central stack of ST is not a serious bottleneck for an aneutronic proton-boron (p-$^{11}$B) fusion reactor. The high beta and compactness of ST make it a potentially attractive candidate for p-$^{11}$B fusion reactor. In addition to the aneutronic environment, p-$^{11}$B fusion has the advantage of abundant and cheap fuel. ENN Science and Technology Development Co., Ltd (ENN) has made the roadmap for p-$^{11}$B fusion based on spherical torus [9].

Although p-$^{11}$B fusion has material and fuel advantages, its critical physics challenges (such as requirements of ultra-high ion temperature and very long energy confinement time, high ratio of ion temperature over electron temperature to reduce the bremsstrahlung and cyclotron radiation, etc.) exceed those of D-T fusion. On the other hand, whether it is used in D-T fusion reactor or p-11B fusion reactor, the ST must solve the issue of high efficiency non-inductive start-up and drive. The tokamak plasma current is required to insure a high plasma confinement capability to restrain transport losses from the core to the edge. The start-up and ramp-up of this current have been commonly driven by a toroidal electric field induced by current changes in a centre solenoid (CS) magnet. This however causes engineering difficulties for ST due to the limited space available in a narrow centre column. To develop a solenoid-free or solenoid-less current drive capability therefore has been an important research endeavour for the STs. On the positive side, removing CS or reducing the size of CS allows additional space to increase the toroidal field (TF), further improving compactness and economy of ST. XuanLong-50 spherical torus (EXL-50) is the first ST device build by ENN group which is the largest private energy supply company in China. The main goal of EXL-50 is to explore the feasibility and effectiveness of fully non-inductive current start-up and drive with electron cyclotron resonance heating (ECRH) without a central solenoid (CS). CS-free ECRH and current drive have been tested in several earlier ST devices (CDX-U [10], LATE [11-15], TST-2 [16-17], MAST [21-22] and QUEST [20-26]). In this paper, we present the non-inductive ECRH current drive experimental results from EXL-50. Not only are the operational parameters of CS-free current drive by ECRH significantly expanded, but some remarkable plasma behaviour is also observed. Moreover, neutral beam heating (NBI), low hybrid current drive (LHCD), and ion cyclotron resonance heating (ICRH) are also tested for short experimental period recently in EXL-50 before the shut-down on June 30 2023. XuanLong-50U (EXL-50U), an upgrade ST, has been designed and will be operated in 2024 in ENN.

The remainder of this paper is organized as follows. The experiment setup in EXL-50 is given in section 2. The main experimental progress and results are described in section 3. Summary and future plans of EXL-50U are presented in section 4.

## 2. EXL-50 SPHERICAL TORUS

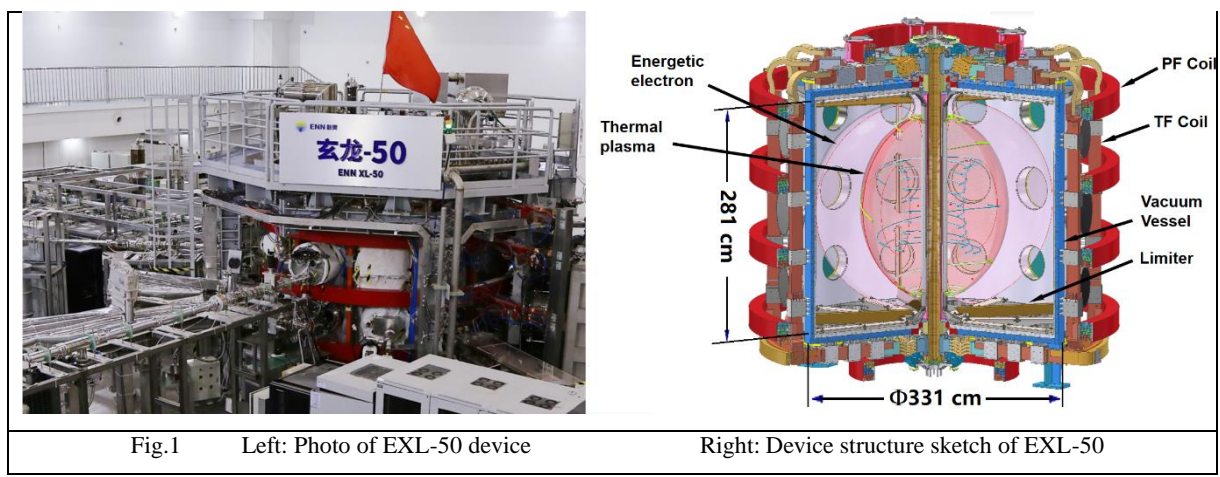

Fig.1    Left: Photo of EXL-50 device            Right: Device structure sketch of EXL-50

The EXL-50 device is the first medium-sized spherical torus (ST) in China. The main machine body of EXL-50 is a cylindrical vacuum vessel, as shown in Fig.1. The engineering design of EXL-50 began in October 2018 and the construction of the device was completed in June 2019. The first plasma discharge was successfully conducted in July 2019. The toroidal field (TF) of EXL-50 is around 0.5 T at major radius of 50 cm. An important characteristic of EXL-50 is that it does not have a central solenoid. Six poloidal field (PF) coils are located outside the vacuum vessel and the TF coil conductors. Inner limiters on the center column and outer limiters on the vessel wall have leading edges at 0.186 and 1.512 m in the major radius, respectively. The inner limiters are made of pure tungsten; the outer limiters are made of tungsten plated copper. There are no other first-wall structure between the plasma and vacuum chamber except for limiters. The main parameters are shown in Table.1.

| Parameter | Value |
| --- | --- |
| Plasma current | 0.2 MA |
| Thermal ions major radius $R_i$ | 58 cm |
| Energetic electron cloud radius | 70 cm |
| Thermal ions aspect ratio (LCFS) | 1.5 |
| Energetic electron cloud aspect ratio | 1.3 |
| Toroidal magnetic field ($R = 0.5$m) | 0.5T |
| Elongation | $\approx 2$ |
| Discharge TF flattop duration | 5s @ 0.5T |
|  | 20s @ 0.3 T |

Table.1 The main parameters of EXL-50

The heating and current drive (H&CD) system is composed of ECRH, ICRF, LHCD and NBI systems. Fig.2 shows the arrangement of H&CD systems on EXL-50. There are total of 6 ECRH systems on EXL-50, 5 of which operated at a frequency of 28GHz (#00 - #04 ECRH in fig.2B), and one operates at a frequency of 50GHz (#05 ECRH in fig.2B). The #00 28GHz ECRH system is equipped with a gyrotron that has a source power of 50 kW and a maximum pulse duration of 30 seconds. This low-power ECRH system, capable of delivering 20 kW of power, is primarily used for pre-ionization and current start-up. The other four 28GHz ECRH systems (#01 - #04 in fig.2B) have a source power of 400 kW and a maximum pulse duration of 5 seconds. One of the 28 ECRH systems is used as a backup. Due to the limitations of transmission system, the actual delivery power of these 28GHz or 50GHz systems is approximately 150kW. The ECRH system serves as the primary heating and current drive (H&CD) power source for EXL-50 and is capable of injecting a maximum power of 600 kW using four gyrotrons during real experiments. The power of ECRH in this paper is the power measured at the matching optical unit (MOU), which is close to the emission power of the gyrotrons. The power delivered from the antenna inside the vacuum vessel is unknown at present due to the lack of monitoring equipment. The LHCD system consists of a 200kW klystron which is typically injected for current drive during flattop phase of ECRH plasmas. The available suppliers only provide klystrons at two frequencies: 2.45 GHz and 3.7 GHz. We chose the 2.45 GHz klystron, which is more suitable for low-magnetic-field ST devices compared to the 3.7 GHz option. The ICRF includes a 100 kW tunable frequency system (3–26 MHz) and a 40 kW system operating at 13.56 MHz, which is used for test of the engineering feasibility at present. A 50 kV NBI with a pulse length of 5 s and a power of 1.5 MW has been designed and installed on EXL-50, and was put into initial test on in the 2023's experimental campaign.

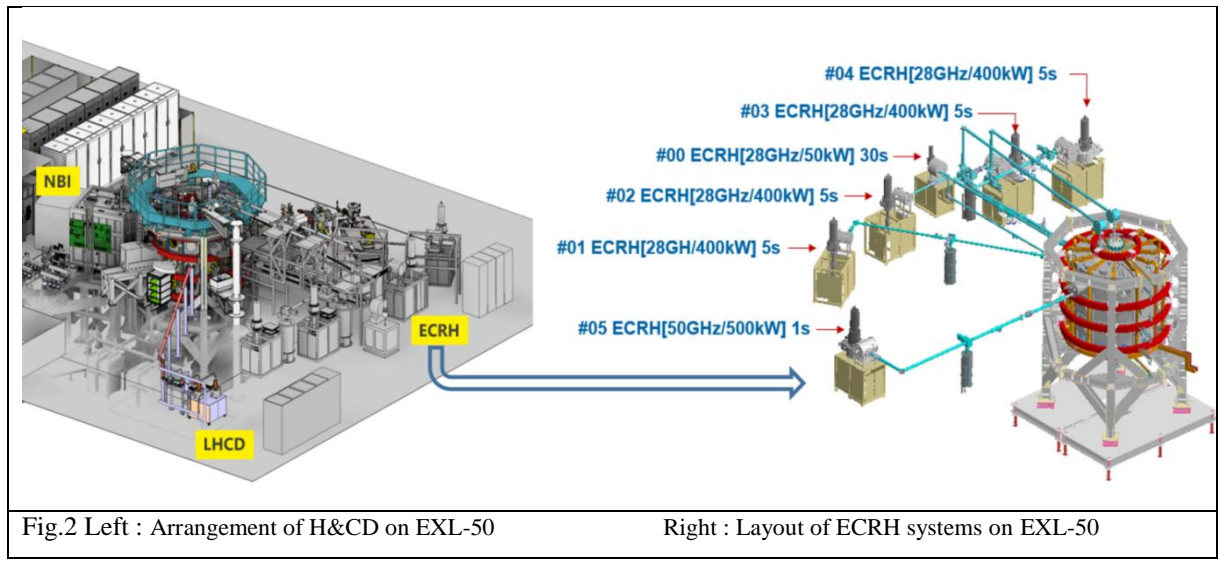

Fig.2 Left : Arrangement of H&CD on EXL-50    Right : Layout of ECRH systems on EXL-50

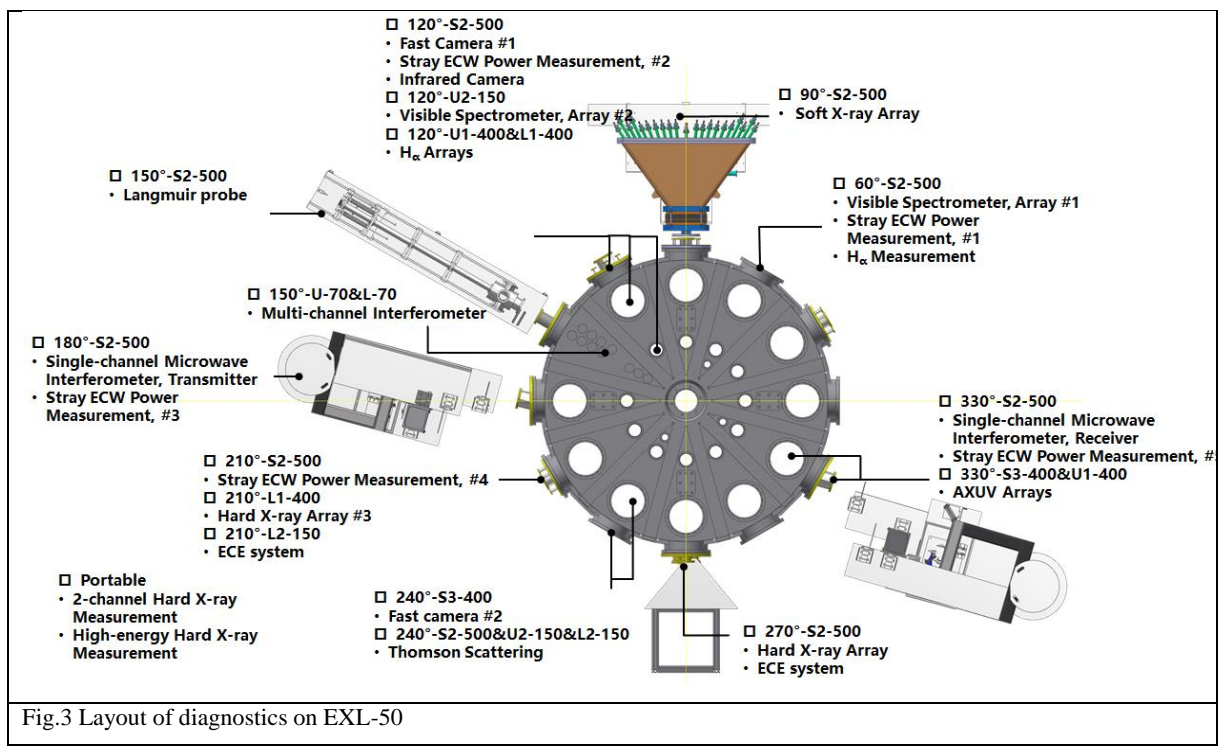

Fig.3 Layout of diagnostics on EXL-50

The diagnostics of the EXL-50U [27-34] are critical for machine safety monitoring, plasma feedback control, and physics analysis. These diagnostic tools include magnetic measurements, visible/infrared cameras, microwave and HCN/FIR interferometers, Thomson scattering diagnostics, Hα and impurity spectroscopy, stray ECW power monitors, ECE measurement systems, AXUV arrays, soft X-ray and hard X-ray arrays, and Langmuir probes. Figure 3 shows the diagnostics layout on EXL-50. The main parameters and functions of these diagnostics are described in Table.2.

| Diagnostics | Measurements | Description |
| --- | --- | --- |
| Fast Camera [27] | Image information | |
| Infrared Camera [27] | Heat load on the first wall | |
| Langmuir Probes | Edge Te and ne | Movable |
| Single-channel Microwave interferometer [28] | Line integrated ne | 140 GHz |
| Multi-channel interferometer [29] | | HCN, 890 GHz, 3-chnnel (Vertical) |
| | | Microwave, 330 GHz, 2-channel (Vertical) |
| Hα array | Line-integrated Hα emission | 656.3 nm |
| | Impurity emission | HeII 468.5 nm<br>OII 441.5 nm<br>CIII 464.7 nm |
| Visible survey spectrometer | Ions temperature | 370-800 nm, R = 19 cm - 85 cm |
| AXUV array [30] | AXUV emission | 16 channel, 1eV to 10keV |
| Thomson scattering systems [31] | Te and ne profile | 1064 nm @ 3J, 15 locations range from Z = -48 cm to 8 cm, R = 70 cm |
| Hard X-ray detector [32] | Hard X-ray emission by energetic electrons | 20-200 keV |
| Soft X-ray detector [33] | Soft X-ray spectra | 1-20 keV |
| Stray ECW power detector | ECW stray power emission | 28 ±0.5 GHz |
| ECE system [34] | ECE emission by energetic electrons | K- and K$\alpha$-band (Horizontal)<br>C-, X- and Ku-band (Vertical) |

Table.2 Parameters and functions of the diagnostics on EXL-50.

Boronization in glow discharges with mixed gas of 30% Diborane ($B_2H_6$) and 70% Hydrogen is the regular wall condition and cleaning method for EXL-50. The investigation of physical properties of magnetic confined hydrogen-boron plasmas is also one key issue for ENN's fusion research project. Several kinds of boron injection system have been developed and installed in EXL-50. The first boron injection system is the gas-puffing system with mixed gas of Diborane ($B_2H_6$) and Hydrogen which is installed in the middle of center stack. The second boron instrument is the boron powder injector which is installed on a top port of EXL-50. The boron powder injector is equipped with two cartridges, allowing for the preloading of boron powder in two different sizes simultaneously. The boron powder used in the experiments on EXL-50 has a particle size of 150 microns. The boron powder injection rate (from 1 mg/s to 10 mg/s) and injection duration (from 50 ms to several seconds) can both be flexibly controlled. The third boron instrument is the boron pellet injector which is installed on a central horizontal port of EXL-50. The diameter and speed of the boron pellet for EXL-50's plasma is 0.5 mm~1 mm and 50 m/s-100 m/s. Both the boron powder and boron pellets are made from boron with a purity of 99.9%. The schematic diagrams of boron powder injector and boron pellet injector are shown in fig.4. These boron injector instruments provide the flexible control for the global boron density and boron injection depth, which are powerful tools for study of hydrogen-boron plasma physics. Moreover, these instruments have been regularly utilized applied for the active real-time wall conditioning in normal plasma discharges on EXL-50.

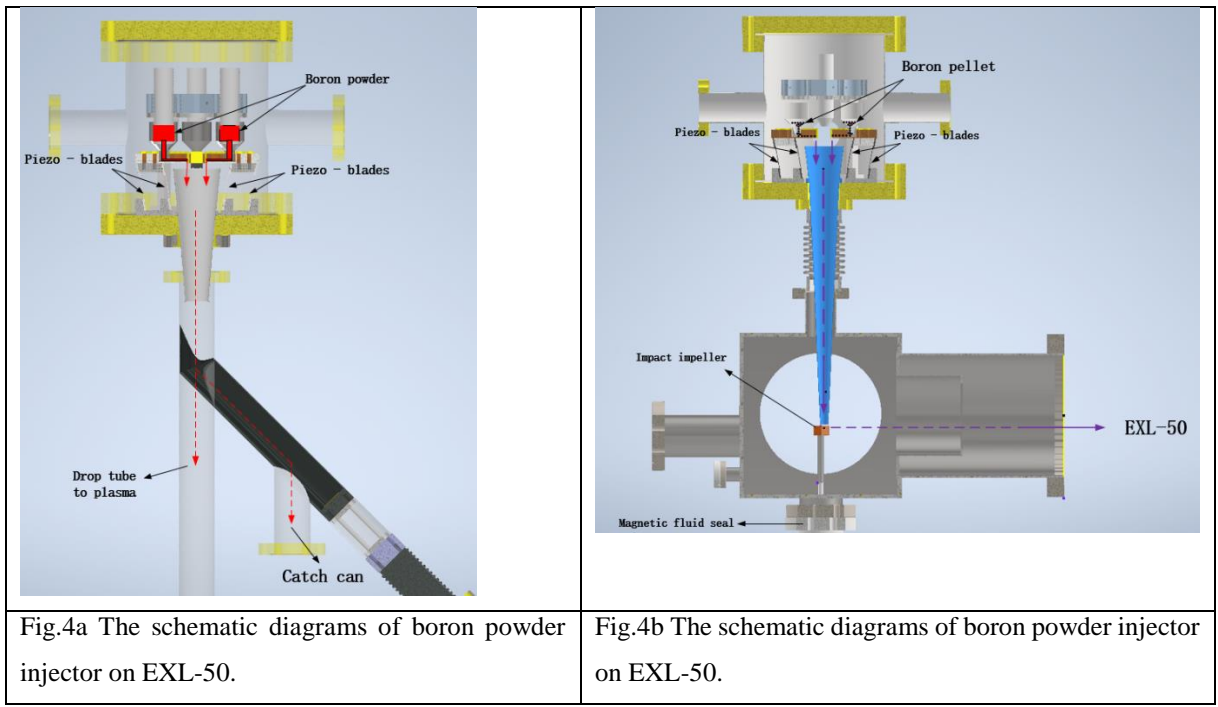

| Fig.4a The schematic diagrams of boron powder injector on EXL-50. | Fig.4b The schematic diagrams of boron powder injector on EXL-50. |
|---|---|

## 3. EXL-50 RESEARCH PROGRESS

The main heating in the discharges in EXL-50 from 2020 to 2022 is 28GHz ECRH [35]. For the 28GHz ECRH experiment phase, plasma currents of 50kA -160 kA are routinely obtained in EXL-50, with the current flattop sustained for up to or beyond 2 s. The average current drive effectiveness on EXL-50 reached around 1 A/W for discharges using single ECRH gyrotron up to 160 kW. The plasma current reaches $I_p > 80$ kA for densities $>5 \times 10^{18}$ m$^{-2}$ using 150 kW ECRH power [35]. Fig.5-7 shows the typical discharges waveforms of 28GHz ECRH plasmas. More detail of 28 GHz ECRH experiment can be seen in ref.35.

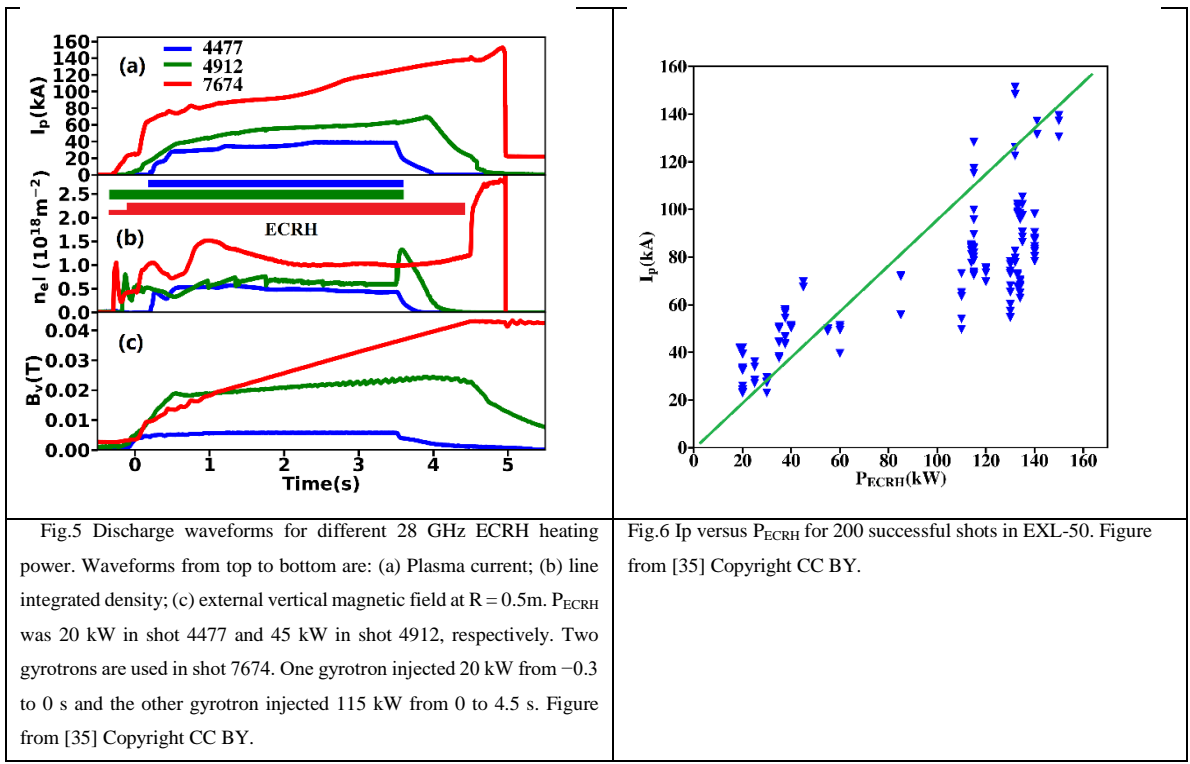

| Fig.5 Discharge waveforms for different 28 GHz ECRH heating power. Waveforms from top to bottom are: (a) Plasma current; (b) line integrated density; (c) external vertical magnetic field at R = 0.5m. P$_{ECRH}$ was 20 kW in shot 4477 and 45 kW in shot 4912, respectively. Two gyrotrons are used in shot 7674. One gyrotron injected 20 kW from −0.3 to 0 s and the other gyrotron injected 115 kW from 0 to 4.5 s. Figure from [35] Copyright CC BY. | Fig.6 Ip versus P$_{ECRH}$ for 200 successful shots in EXL-50. Figure from [35] Copyright CC BY. |
|---|---|

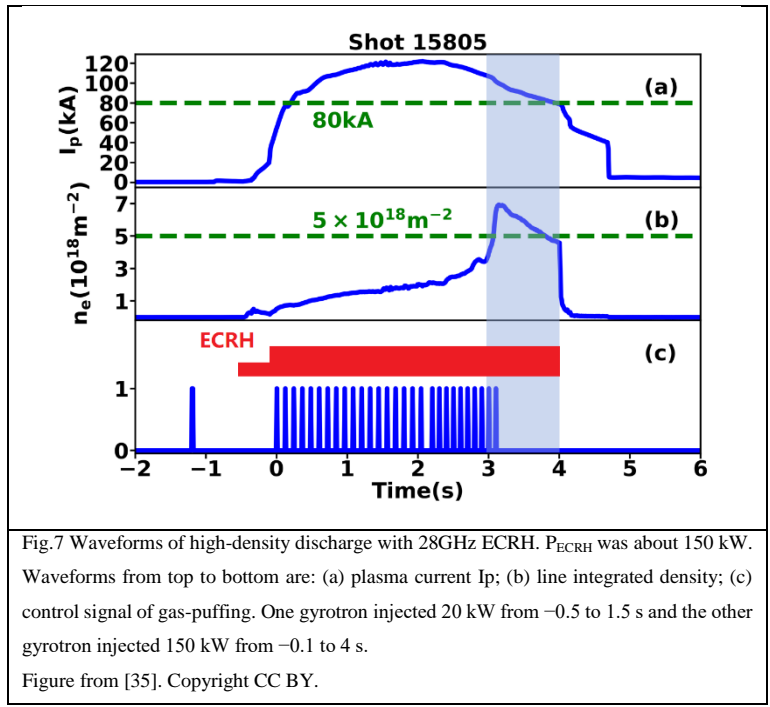

Fig.7 Waveforms of high-density discharge with 28GHz ECRH. $P_{ECRH}$ was about 150 kW. Waveforms from top to bottom are: (a) plasma current Ip; (b) line integrated density; (c) control signal of gas-puffing. One gyrotron injected 20 kW from −0.5 to 1.5 s and the other gyrotron injected 150 kW from −0.1 to 4 s.
Figure from [35]. Copyright CC BY.

A new 50GHz gyrotron was put into use in 2023's experimental campaign. Improved plasma performance was obtained. For the typical 50GHz ECRH discharge experiments, the 28GHz ECRH is still applied to plasma current start-up and ramp-up. As shown in Fig.8, 28GHz ECW turned off at 1.0s and then 50GHz ECRH was injected. The maximum plasma current can reach 180 kA after the injection of 50GHz ECRH. On the other hand, the density increase during 50GHz ECRH phase is very clear and remarkable in discharges with moderate 120kA plasma current, as shown in Fig.9. The reconstructed density profile based on the core line integrated density from HCN interferometer data and the edge local density from Thomson scattering is shown in the Fig.10. It can be seen that the reconstructed core density of 50GHz ECRH plasmas can exceed $1\times 10^{19}$ m$^{-3}$, if parabolic profiles are assumed. The waveforms of the other typical discharge are also shown in Fig.11. For such discharges, two 28GHz ECRH gyrotrons turn on during the injection of 50GHz. The line integrated density during the overlap phase of 28GHz & 50GHz in Fig.11 is about twice as that in the 50GHz alone phase in Fig.9. There is no valid Thomson scattering data for these high density discharges and reconstructed density profile cannot be obtained. Nevertheless, the core plasma density during 28GHz+50GHz phase might have reached $2\times 10^{19}$ m$^{-3}$ for if the shape of density profile were similar for the discharges in Fig.9 and Fig.11.

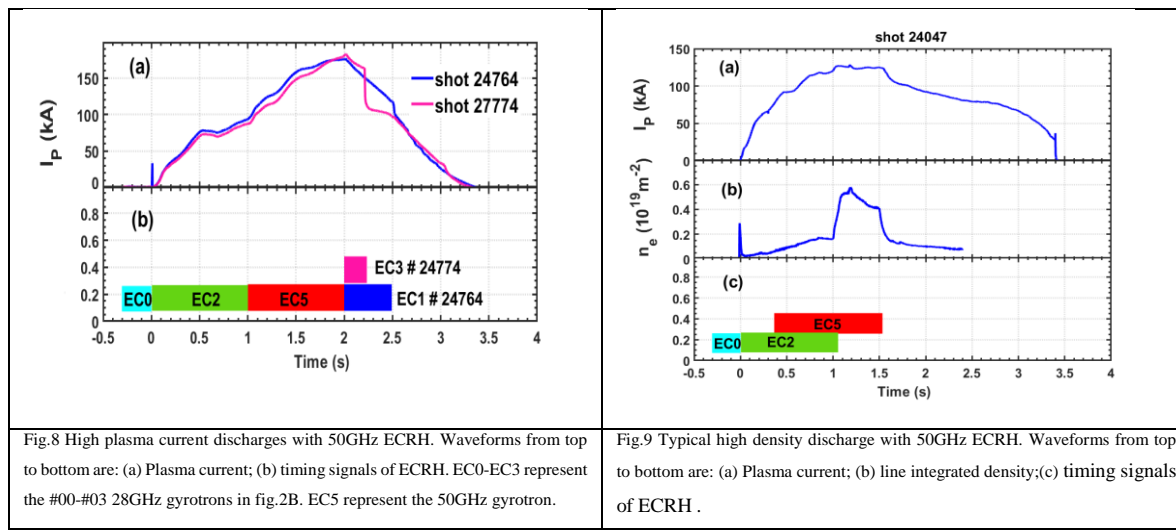

Fig.8 High plasma current discharges with 50GHz ECRH. Waveforms from top to bottom are: (a) Plasma current; (b) timing signals of ECRH. EC0-EC3 represent the #00-#03 28GHz gyrotrons in fig.2B. EC5 represent the 50GHz gyrotron.

Fig.9 Typical high density discharge with 50GHz ECRH. Waveforms from top to bottom are: (a) Plasma current; (b) line integrated density;(c) timing signals of ECRH .

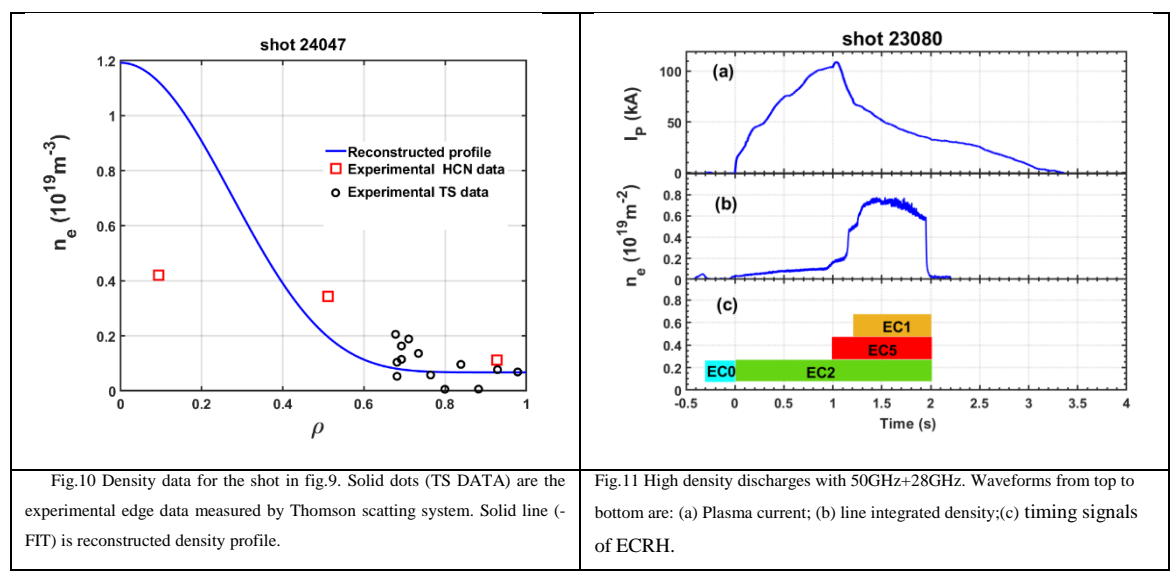

Fig.10 Density data for the shot in fig.9. Solid dots (TS DATA) are the experimental edge data measured by Thomson scatting system. Solid line (-FIT) is reconstructed density profile.

Fig.11 High density discharges with 50GHz+28GHz. Waveforms from top to bottom are: (a) Plasma current; (b) line integrated density;(c) timing signals of ECRH.

The energetic electrons play a unique and important role in EXL-50's plasmas. EXL-50 experimental results indicate that the plasma current is mainly carried by the energetic electrons [27, 35, 36]. The metal wall of the vacuum vessel effecting multiple reflections and absorption at high multi-harmonic resonances increases the high efficiency of acceleration of the energetic electrons [35]. The simulation with CQL3D and GENRAY shows that the multi-pass absorption and multi-harmonic of ECW can greatly improve the current drive efficiency [37]. On the other hand, the asymmetric distribution of the energetic electrons in velocity space based on orbit analysis in a multi-fluid equilibrium [38] is another key feature of the very high current drive effectiveness observed in EXL-50. The modelling and simulation based on the orbit mechanism of energetic electrons also can be seen in [39]. It should be noted that the role of induction effective in the CS-free ECRH-driven current remains a question. The changes in the PF coils still can induce toroidal electron field. In [35], some dedicated experiments and analysis are provided to clarify the effect of inductive current. Here, the new experiments results related to this issue are also shown in Fig.12. The current of the all PF coils in the discharge in fig.12 kept constant from -1s to 3s. ECRH is injected from 0s and then plasma current is start-up and ramp-up to flattop phase. These dedicate discharges in fig.12 clearly demonstrate that the inductive plasma current is negligible. Fig.13 shows the waveforms of the other dedicated shots. The toroidal injection angel and power of ECRH are same for the two shots in fig.13. It clear shows the direction of plasma current $I_p$ depends on the direction of external vertical field $B_v$ which determines the asymmetric distribution of energy electrons in parallel direction. The experiment results in fig.13 clearly confirm the asymmetric confinement mechanism of energetic electrons for the non-inductive current drive.

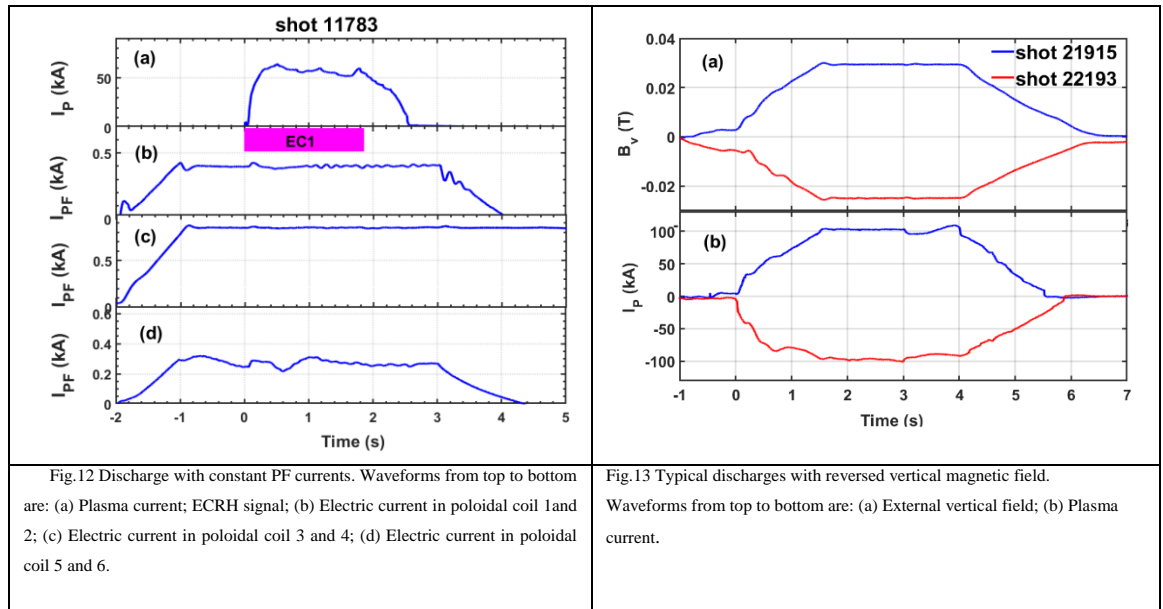

Fig.12 Discharge with constant PF currents. Waveforms from top to bottom are: (a) Plasma current; ECRH signal; (b) Electric current in poloidal coil 1 and 2; (c) Electric current in poloidal coil 3 and 4; (d) Electric current in poloidal coil 5 and 6.

Fig.13 Typical discharges with reversed vertical magnetic field. Waveforms from top to bottom are: (a) External vertical field; (b) Plasma current.

Fig.14 shows the relation between plasma current and density for the typical dedicated and well-controlled shots in EXL-50. It can be seen that the decreasing trend of plasma current density clearly deviates from the linear dependence ($I_p \propto 1/n_e$) by traditional ECCD theory. Moreover, the normalized current drive efficiency increases with the line averaged density, as shown in fig.15. The toroidal field for the data in fig.14 and fig.15 is same (~0.5T @ R = 50 cm) for 28GHz and 50GHz ECRH. Fig.14 and Fig.15 also clearly demonstrate that high frequency ECRH is helpful to increase plasm current and current drive efficiency.

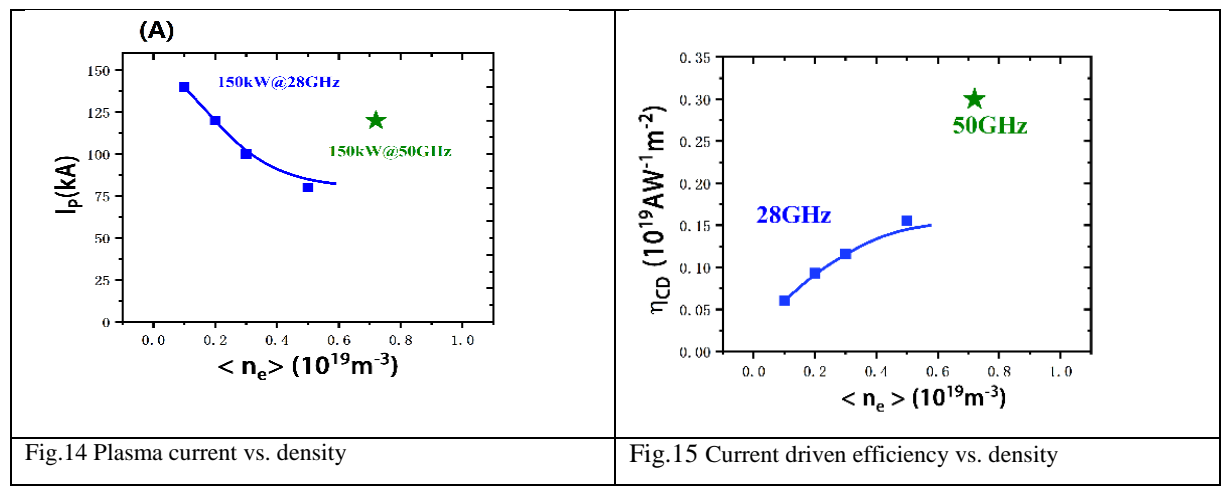

Fig.14 Plasma current vs. density

Fig.15 Current driven efficiency vs. density

The plasma current can be driven by LHCD, as shown in figure 16. The background plasma parameters in the two shots in fig.16 are same. The background plasma and current in the two discharges are produced and sustained by 150 kW ECRH. 100 kW LHW is injected between 2.5s and 3.5s in shot 19704. The plasma current increases by about 20 kA during LHCD phase. The plasma density also increased after the injection of LHW.

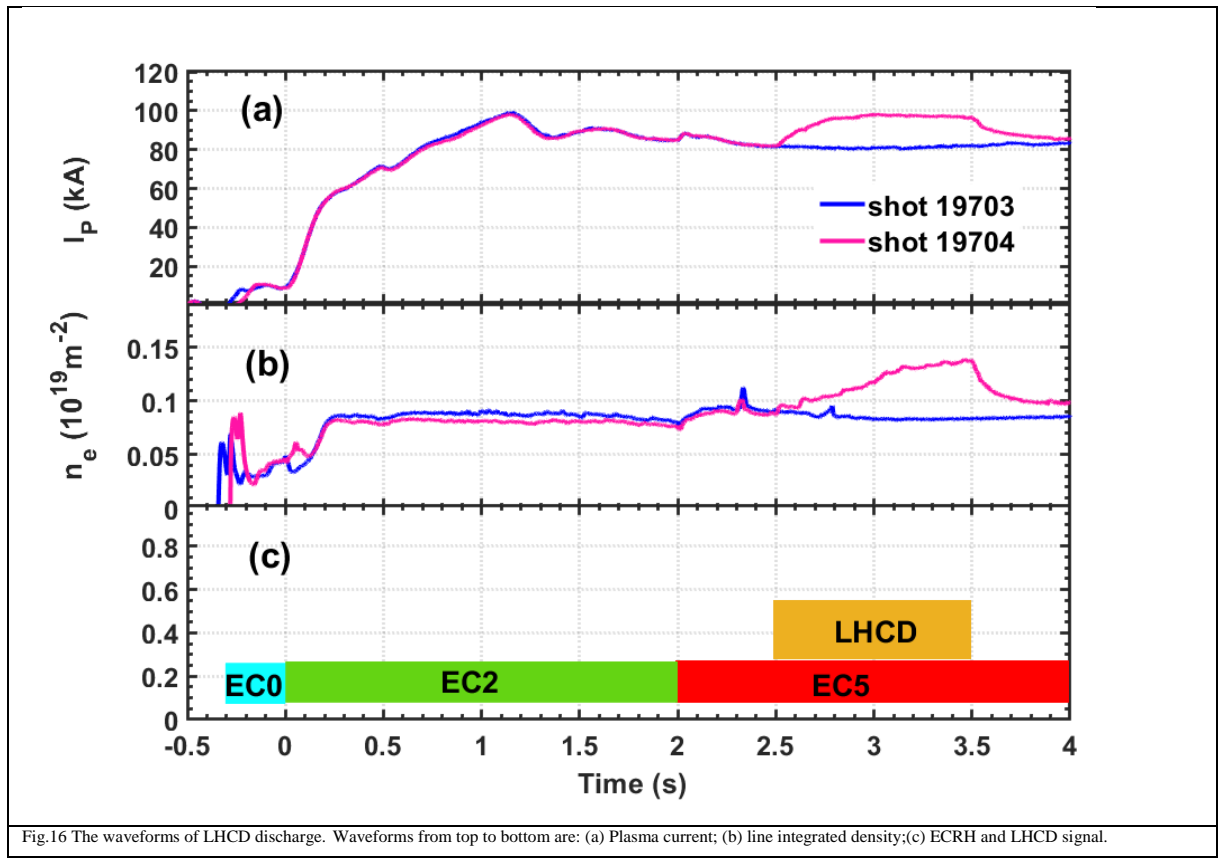

Fig.16 The waveforms of LHCD discharge. Waveforms from top to bottom are: (a) Plasma current; (b) line integrated density;(c) ECRH and LHCD signal.

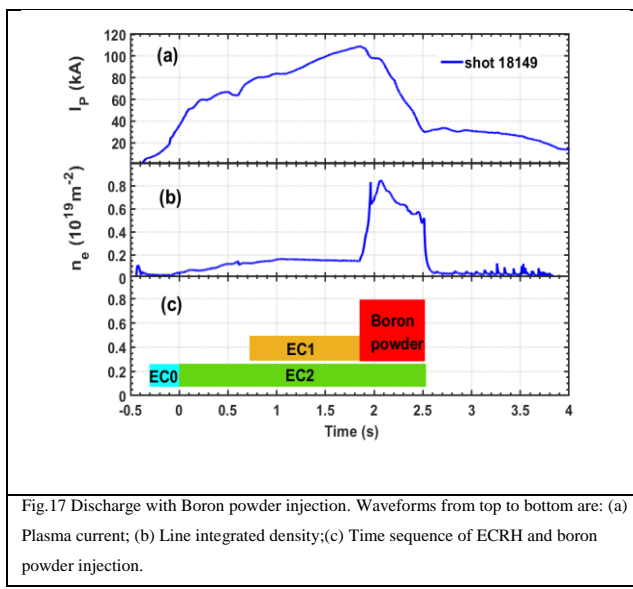

Fig.17 Discharge with Boron powder injection. Waveforms from top to bottom are: (a) Plasma current; (b) Line integrated density;(c) Time sequence of ECRH and boron powder injection.

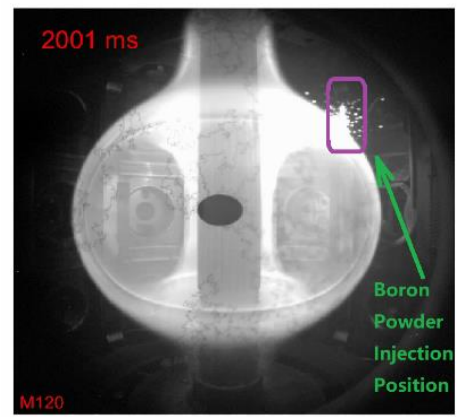

*Fig.18 Plasma image from fast camera during boron powder injection*

The boron power injection has been used to real-time wall conditioning; the power was injected from the top of device at major radius $R = 70$ cm. The typical discharge waveforms of powder injection are shown in fig.17. The boron power is injected from 1.9 s to 2.5 s. It can be seen in the photos in fig.18 that boron power ablated at the region outside of LCFS. The injection of boron power can suppress plasma-wall interaction and edge turbulence. A clear plasma boundary was observed from the fast CCD cameras, as shown in fig.18, and the plasma boundary moves further away from the vacuum wall. The plasma density increased significantly from $1.5 \times 10^{18}$ m$^{-2}$ to $8.0 \times 10^{18}$ m$^{-2}$, i.e., 430%, and the results can be found in figure 17.

The experiment results of boron pellet injection are shown in Fig.19. Compared to boron powder, boron pellet injection can provide deeper fuelling. The injection depth of pellet estimated from the fast camera is around 5cm-10cm. The main plasma parameters and setting for the two shots are identical. The pellet is injected at 1.3 s in shot 25913 and the density jumps after the injection of pellet and decay slowly. After the density recovery to the level before pellet injection. Gas puffing is applied to increase density at around 1.5s in the two shots in fig.19. It can be seen than the density after 1.5 s for pellet shot 25913 is obviously higher than the shot 25912 without pellet injection. This phenomenon indicate that injection of pellet can also improve the efficiency of gas fuelling. It should be point out that the temperature could be significantly decreased due to the injection of the Boron powder or pellets. These experiments demonstrate the capability and reliability of boron injection hardware system. The effects of boron injection on the plasma confinement and transport will be systematically and comprehensively on the EXL-50U, which is the upgraded device of EXL-50.

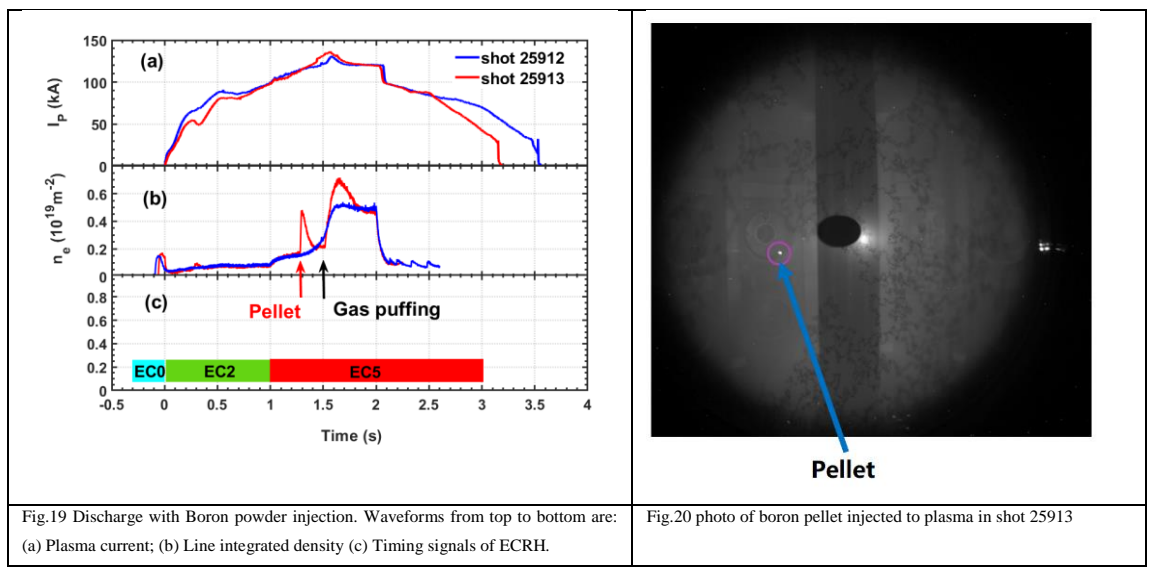

Fig.19 Discharge with Boron powder injection. Waveforms from top to bottom are: (a) Plasma current; (b) Line integrated density (c) Timing signals of ECRH.

Fig.20 photo of boron pellet injected to plasma in shot 25913

## 4. SUMMARY AND FUTURE PLAN

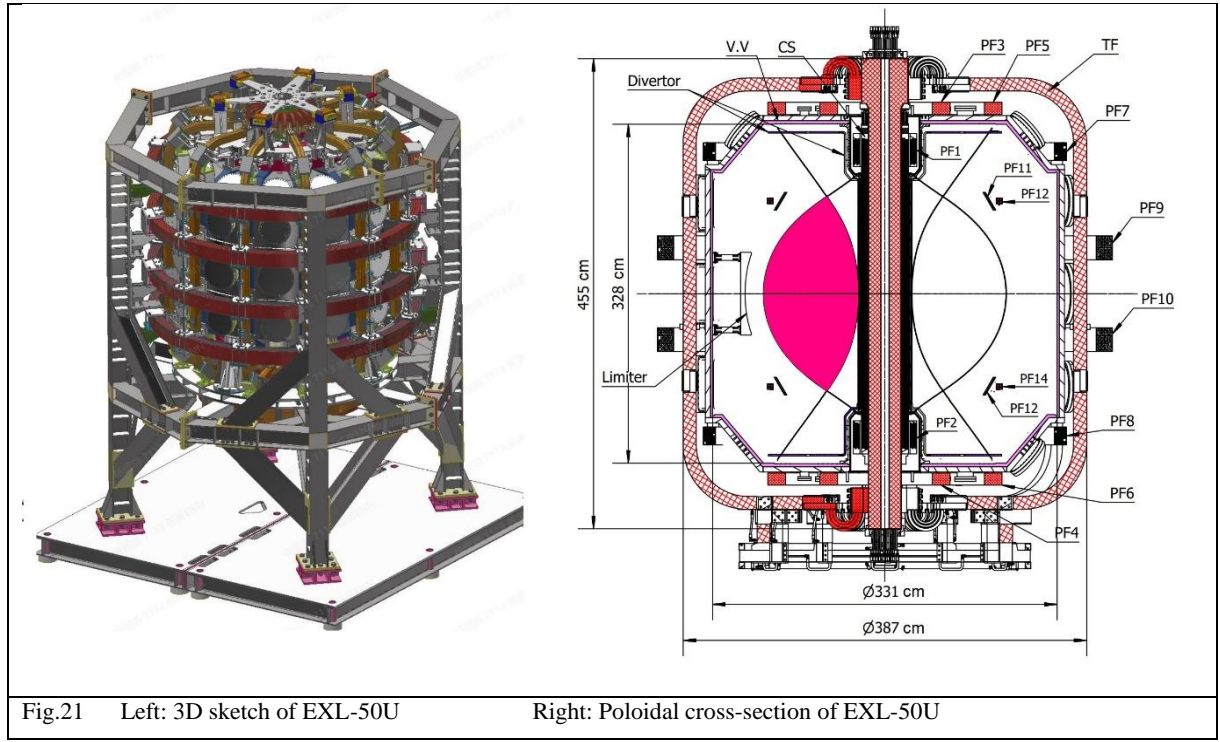

Fig.21　Left: 3D sketch of EXL-50U　　Right: Poloidal cross-section of EXL-50U

| Parameters | Value |
|---|---|
| Plasma current | 0.5MA |
| Major radius | 60 – 80 cm |
| Toroidal magnetic field ($R = 60$ cm) | 1.2T |
| Aspect ratio | 1.4-1.85 |
| Elongation | 2 |
| Discharge TF flattop duration | 2s |
| NBI | 1.5MW/50kV/5s |
| | 1MW/25kV/2s |
| ECRH | 2×0.4MW/28GHz/5s |
| | 2×0.4MW/50GHz/1s |
| | 1×0.5MW/80GHz/1s |
| ICRF | 2MW/25MHz~40MHz/1s |
| LHCD | 2×0.2MW/2.45GHz/CW |

Table.3 Main parameters of EXL-50U

CS-free ECRH experiment has been systemically investigated on EXL-50. The averaged current drive effectiveness on EXL-50 is around 1 A/W, which is around twice that (0.5 A/W) of previous similar experiments [22]. The plasma current routinely researches 180 kA, which is also higher than the maximum value (~90 kA) of a previous CS-free ECRH experiment [25]. Both experimental results and theory & modelling indicate that the plasma current in EXL-50 is carried by the energetic electrons. There are two possible physics mechanisms for the non-inductive current drive experiment in EXL-50. One is the multiple reflections & absorption with high multi-harmonic resonances of ECRH which increases the high efficiency of acceleration of the energetic electrons. The other is the asymmetric distribution of the energetic electrons in velocity space based on orbit analysis. The two mechanisms are not contradictory, but can coexist and work synergistically. The experimental results of EXL-50 also show the positive effects of ECRH frequency on the current drive efficiency.

On the other hand, EXL-50 will be replaced by a new spherical torus which is named as EXL-50U. Although the machine size of EXL-50U is similar as EXL-50, the TF field of EXL-50U will increase to 1.2T @ $R$ = 60cm. CS coil and more PF coils will be installed on EXL-50U to improve plasma shaping and current control. The main body of EXL-50U will be assembled at the same position of EXL-50. In addition to the current CD&H systems on EXL-50 which will continue to work and test on EXL-50U, one new 20kV NBI and some new higher frequency ECRH and high power ICRF are also planned to be installed to obtain high parameter plasmas. The key physics issues of EXL-50 are as following: scenarios development for stable high density hot ion mode for ST ($T_{i0}$ = 3keV~ 5keV, $n_{e0}$ = 5~8×10$^{19}$ m$^{-3}$), investigation of energy confinement scaling for wide range scan of aspect ratio (1.4~1.8) and $B_t$ (0.5T~1.2T), high density non-inductive current drive, Proton-Boron plasma physics. The design of EXL-50U has been completed and main components are already being manufactured. The machine assembly of EXL-50U is scheduled to be completed in the end of 2023. The first plasma of EXL-50U is expected to be obtained in the beginning of 2024.

**REFERECES**

[1] Peng Y.-K.M. and Strickler D.J., Features of spherical torus plasmas, Nucl. Fusion **26** (1986) 769–77

[2] Peng Y.-K.M. , The physics of spherical torus plasmas Phys. Plasmas **7** (2000) 1681–9

[3] Sykes A., et al., High-performance of the START spherical tokamak, Plasma Phys. Control. Fusion **39** (1997) B247–60

[4] Synakowski E.J., et al., he national spherical torus experiment (NSTX) research programme and progress towards high beta, long pulse operating scenarios, Nucl. Fusion **43** (2003) 1653–64

[5] Buttery R.J., et al, Stability at high performance in the MAST spherical tokamak, Nucl. Fusion **44** (2004)1027–35

[6] Kurskiev G. S., et al., Energy confinement in the spherical tokamak Globus-M2 with a toroidal magnetic field reaching 0.8 T, Nuclear Fusion **62** (2022) 016011

[7] Kurskiev G. S., et al., The first observation of the hot ion mode at the Globus-M2 spherical tokamak, Nuclear Fusion **62** (2022) 104002

[8] McNamara S. A. M., et al., Achievement of ion temperatures in excess of 100 million degrees Kelvin in the compact high-field spherical tokamak ST40, Nucl. Fusion **63** (2003) 054002

[9] LIU M.S., et al., ENN's Roadmap for Proton-Boron Fusion Based on Spherical Torus, IAEA 29th Fusion Energy Conference, London, UK, Oct.16-21, 2023

[10] FOREST C.B., HWANG Y.S., ONO M. and DARROW D.S., Internally generated currents in a small-aspect-ratio tokamak geometry Phys. Rev. Lett. **68** (1992) 3559–62

[11] MAEKAWA T. et al., Formation of spherical tokamak equilibria by ECH in the LATE device Nucl. Fusion **45** (2005) 1439–45

[12] YOSHINAG T., et al., Spontaneous formation of closed-field torus equilibrium via current jump observed in an electron-cyclotron-heated plasma Phys. Rev. Lett. **96** (2006)125005

[13] UCHIDA M., et al., Rapid current ramp-up by cyclotron-driving electrons beyond runaway velocity Phys. Rev. Lett. **104** (2012) 065001

[14] KURODA K., et al., Shift in principal equilibrium current from a vertical to a toroidal one towards the initiation of a closed flux surface in ECR plasmas in the LATE device Plasma Phys. Control. Fusion **58** (2016) 025013

[15] TANAKA H., 2018Electron Bernstein wave heating and current drive with multi-electron cyclotron resonances during non-inductive start-up on LATE 2018 IAEA Fusion Energy Conf. (Gandhinagar) EX/P3-19 (Ahmedabad, India) (https://nucleus.iaea.org/sites/fusionportal/Shared%20Documents/ FEC%202018/fec2018-preprints/preprint0078.pdf)

[16] EJIRI A. et al., Non-inductive plasma current start-up by EC and RF power in the TST-2 spherical tokamak Nucl. Fusion **49** (2009) 065010

[17] TAKASE Y. et al., Non-inductive plasma initiation and plasma current ramp-up on the TST-2 spherical tokamak Nucl. Fusion **53** (2013) 063006

[18] SHEVCHENKO V.F., et al, Electron Bernstein wave assisted plasma current startup in MAST Nucl. Fusion **50** (2010) 022004

[19] SHEVCHENKO V.F. et al., Long pulse EBW start-up experiments in MAST EPJ Web Conf. **87** (2015) 02007

[20] HANADA K. et al., Non-inductive start up of QUEST plasma by RF power Plasma Sci. Technol. 13 (2011) 307–11

[21] ISHIGURO M. et al., Non-inductive current start-up assisted by energetic electrons in Q-shu University experiment with steady-state spherical tokamak Phys. Plasmas **19** (2012) 062508

[22] TASHIMA S. et al., Role of energetic electrons during current ramp-up and production of high poloidal beta plasma in non-inductive current drive on QUEST Nucl. Fusion **54** (2014) 023010

[23] IDEI H. et al., Fully non-inductive second harmonic electron cyclotron plasma ramp-up in the QUEST spherical tokamak Nucl. Fusion **57** (2017)126045

[24] IDEI H. et al., Fully non-inductive 2nd harmonic electron cyclotron current ramp-up with polarized focusing-beam in the quest spherical tokamak 2018 IAEA Fusion Energy Conf. (Gandhinagar) EX/P3-21 Ahmedabad, India) (https://nucleus.iaea.org/sites/fusionportal/Shared%20Documents/ FEC%202018/fec2018-preprints/preprint0175.pdf)

[25] IDEI H. et al., Electron heating of over-dense plasma with dual-frequency electron cyclotron waves in fully non-inductive plasma ramp-up on the QUEST spherical tokamak Nucl. Fusion **60** (2020) 016030

[26] ONCHI T. et al 2021 Non-inductive plasma current ramp-up through oblique injection of harmonic electron cyclotron waves on the QUEST spherical tokamak Phys. Plasmas **28** (2021) 022505

[27] GUO D. et al, Experimental study of the characteristics of energetic electrons outside LCFS in EXL-50 spherical torus, Plasma Phys. Control. Fusion **64** (2022) 055009

[28] LI S.J. et al., A quasi-optical microwave interferometer for the XuanLong-50 experiment, J. Instrum., **16** (2021) T08011

[29] XIE J.X., et al., Development of a combined interferometer using millimeter wave solid state source and a far infrared laser on ENN's XuanLong-50 (EXL-50), Plasmas Sci. Technol., **24** (2022) 064004

[30] DAI L. L., et al., Time resolved absolute extreme ultraviolet radiation measurement on the ENN XuanLong-50 spherical tokamak, Rev. Sci. Instrum., **92** (2021) 083507

[31] LI H. Y. et al., Thomson scattering diagnostic system for the XuanLong-50 experiment, Rev. Sci. Instrum., **93** (2022) 053504

[32] CHENG S. K. et al, Tangential hard x-ray diagnostic array on the EXL-50 spherical tokamak, Rev. Sci. Instrum., **92** (2021) 043513

[33] HUANG X. L., et al., Toroidal soft x-ray array on the EXL-50 spherical tokamak, Rev. Sci. Instrum. **92** (2021) 053501

[34] WANG Y. M., et al., Design of the electron cyclotron emission diagnostic on EXL-50 spherical torus, Plasmas Sci. Technol., submitted (2023)

[35] SHI Y.J., et al., Solenoid-free current drive via ECRH in EXL-50 spherical torus plasmas, Nucl. Fusion **62** (2022) 086047

[36] WANG M.Y., et al, Experimental study of non-inductive current start-up using electron cyclotron wave on EXL-50 spherical torus, Plasma Phys. Control. Fusion **64**(2022)075006

[37] DEBABRATA B., et al, Investigation of the effectiveness of 'multi-harmonic' electron cyclotron current drive in the non-inductive EXL-50 ST, Journal of Physics: Conference Series **2397** (2022) 012011

[38] ISHIDA A., PENG Y. K. M. and LIU W. J., Four-fluid axisymmetric plasma equilibrium model including relativistic electrons and computational method and results, Phys. Plasmas, **28**(2021)032503

[39] MAEKAWA T. PENG Y. K. M. and LIU W. J., Particle orbit description of cyclotron-driven current-carrying energetic electrons in the EXL-50 spherical torus, Nucl. Fusion **63** (2023) 076014